\documentclass{amsart}

\usepackage{amssymb}
\usepackage{graphicx,color}
\usepackage[round]{natbib}
\usepackage{mathptmx}
\usepackage{microtype}
\usepackage{enumitem}
\usepackage{ulem}
\usepackage{hyperref}
\usepackage{url}

\newtheorem{theorem}{Theorem}{}
{}
\newtheorem{definition}{Definition}{}
\newtheorem{proposition}{Proposition}{}
{}
{}
\newtheorem{remark}{Remark}{}
\newtheorem{example}{Example}{}

\newtheoremstyle{appendix}{}{}{\itshape}{}{\bfseries}{.}{.5em}{#1 \thmnote{#3}}
\theoremstyle{appendix}
\newtheorem*{atheorem}{Theorem}
\usepackage{setspace}
\def\eps	{\varepsilon}
\def\R	{\mathbb{R}}
\def\cG	{\mathcal{G}}
\def\cS	{\mathcal{S}}
\def\C	{\mathcal{C}}
\def\Q	{\mathcal{Q}}
\def\G	{\mathbf{G}}
\def\A	{\mathbf{A}}
\def\Y	{\mathbf{Y}}
\def\Z	{\mathbf{Z}}

\renewcommand{\Pr}[1]	{\mathbb{P}\left( #1 \right)}
\newcommand{\Ex}[1]	{\mathbb{E}\left( #1 \right)}

\def\Var	{\operatorname{var}}

\begin{document}
\title{Non Parametric Statistics of Dynamic Networks with distinguishable nodes}

\author{Daniel Fraiman$^{1,2}$}
\address{$^1$Departamento de Matem\' atica y Ciencias, Universidad de San Andr\' es, Buenos Aires, Argentina.}
\address{$^2$Consejo Nacional de Investigaciones Cient\'ificas y Tecnol\'ogicas (CONICET), Argentina.}
\email[Corresponding author]{dfraiman@udesa.edu.ar}

\author{Nicolas Fraiman$^3$} 
\address{$^3$Department of Statistics and Operations Research, University of North Carolina, Chapel Hill, USA.}
\email{fraiman@email.unc.edu}

\author{Ricardo Fraiman$^4$}
\address{$^4$Centro de Matem\'atica, Facultad de Ciencias, Universidad de la Rep\'ublica, Uruguay.} 
\email{rfraiman@cmat.edu.uy}

\begin{abstract}
The study of random graphs and networks had an explosive development in the last couple of decades. Meanwhile, techniques for the statistical analysis of sequences of networks were less developed. In this paper we focus on networks sequences with a fixed number of labeled nodes and study some statistical problems in a nonparametric framework. We introduce natural notions of center and a depth function for networks that evolve in time. We develop several statistical techniques including testing, supervised and unsupervised classification, and some notions of principal component sets in the space of networks. Some examples and asymptotic results are given, as well as two real data examples.
\end{abstract}

\thanks{This article was produced as part of the activities of FAPESP Center for Neuromathematics (grant\#2013/07699-0, S.Paulo Research Foundation).}
\thanks{The final publication is available at Springer via http://dx.doi.org/10.1007/s11749-017-0524-8}
\maketitle

\section{Introduction}
%==============================
In the last fifteen years, applications of random networks have expanded beyond mathematics, physics, and computer science. Many climatologists, neuroscientists, biologists, sociologists, and economists are becoming more interested in the subject. Networks provide useful representations for many experimental and artificial phenomena.

A large number of research lines in random networks are developed each year. Most of these developments can be classified in one (or both) of the following two categories: probabilistic modelling or statistical analysis. In the former, the focus is on developing new models or deepen the knowledge of an existing model capable of reproducing some network properties. Some important results among these lines include the existence of stationary measures in dynamic models~\citep{watts} or static but growing in size~\citep{barabasi,dani}, characterizations of thresholds for giant components and connectivity~\citep{bollobas,nico}, and analysis of the spread of epidemics over fixed networks. In the latter category, the main goal is to develop mathematical tools for obtaining a precise characterization or description of real networks. Examples of this category include the development of modularity measures~\citep{newman1,bickel}, network motifs, community detection~\citep{newman2, com10,com11}, spectral network analysis~\citep{com9,com6} among others. Our contribution belong to this category of statistical characterization. The literature concentrates mostly on \textit{individual} network analysis. However, we are rather interested in studying a \textit{set} of networks. We propose a new way to describe, classify, perform principal components and testing  for \textit{dependent sequences} of networks with \textit{distinguible nodes}. 

Recently, the study of a sequence of networks has gained more attention. Most of the empirical and theoretical work comes from social media and communication network analysis. Several approaches have been developed to study these dynamical networks, including techniques for networks visualization~\citep{time8}, as well as techniques to describe temporal changes in local and global network properties~\citep{time7,time4,time3,time0}. See~\citet{time5} for a recent review. In regard to modelling, there are proposals which adapt the stochastic block model for evolving networks \citep{block1,block2}. There is a great amount of literature presenting empirical data about time dependent networks \citep[see for example][]{time1,time2, time6}. Nevertheless, as far as we know, the statistical inference of time evolving network has received little attention. A relevant reference in this direction is~\citet{timevarying} where they estimate parametric time varying networks. In contrast, one of the main interesting aspects of our proposal is being non-parametric. In this paper, we discuss how some multivariate statistical methods can be adapted to analyze a random sample of networks or the stochastic dynamics of an evolving network with distinguishable nodes.

In many networks, such like those modelling brain connections, financial markets, the internet, or protein interactions, the label that identifies each node appears naturally and is relevant. For example, a node can represent a particular region of the brain with its own characteristics. Despite this fact the theory of random networks is dominated by models where node labels are not important for the kind of properties studied. In this work we consider the space of networks where each node is distinguishable. In this case, the mathematical tools are simpler mainly because we do not have to deal with identifying isomorphic networks. Specifically, we address the following questions:
\begin{enumerate}
\item[(a)] Given a random network $\G$.
\begin{enumerate}
\item[1.] How to define measures of centrality and variability?
\item[2.] How to define a depth function in the space of networks?
\end{enumerate}
\item[(b)] Given a sample of networks $G_1, G_2\ldots, G_\ell$.
\begin{enumerate}
\item[1.] How to calculate their empirical measures of centrality and variability?
\item[2.] How to perform hypothesis testing for the one and the two sample problem?
\item[3.] How to define a notion of principal components?
\item[4.] How to perform supervised and unsupervised classification?
\end{enumerate}
\end{enumerate}
This short list includes several of the most commonly used techniques in applications in statistics. Our approach provides a way to deal with statistical problems using only the metric structure of the space where the random elements are defined.

We present some answers based on the definition of a depth function, which is defined by a distance in the space of networks, that has the nice property of determining the distribution of a random network. We develop some statistical analysis tools based on it, and show that many standard problems in multivariate analysis can be easily adapted to our framework. 

One major issue when working with network data is the implementation of the proposed techniques for large scale networks. For all the problems addressed in the manuscript we find explicit algorithms which allow to implement our methods with a computational complexity of order $n^2$ (in fact, linear in the number of links), where $n$ is the number of nodes, so we can deal with large scale networks and big datasets. We show several results regarding consistency and asymptotic distribution, and two real data applications. We also exhibit simple and explicit formulae to calculate depth, center and principal components. All the results we describe can be easily adapted for random Boolean functions $f :\{0, 1\}^n \rightarrow \{0, 1\}$ if we forget about the graph representation via the adjacency matrix. However, we are interested in the graph structure obtained from the links between nodes. All proofs are given in the Appendix.

\section{Probability and Graph Theory framework}
%==============================

A network, denoted by $G = (V, E)$, is described by a set $V$ of nodes (vertices) and a set $E \subset V \times V$ of links (edges) between them. In what follows, we consider families of networks defined over the same fixed finite set of $n$ nodes and $m=n(n-1)/2$ stands for the total number of possible links. Then, a network is completely described by its adjacency matrix $A \in \{0,1\} ^{n \times n}$, where $A_{ij}=1$ if and only if the link $(i,j) \in E$.

\subsection{Metrics on the space of networks}
%===============

Describing the similarity of two networks is a relevant question that arised more than 40 years ago \citep[see][for a recent review]{kernels2}. The problem of inexact graph matching considers graphs with indistinguible nodes and/or different number of nodes. In this context, graphs kernels~\citep{kernels,kernels3} and the edit distance~\citep{gao} have been shown to be a good way of extracting information of similarities between graphs. 
The graph edit distance is one of the most flexible distance measures that have been proposed.

For unlabeled networks, the graph edit distance between two networks is defined as the cost of the least expensive sequence of edit operations that are needed to make one network isomorphic to  another one. The edit operations may include insertion, deletion or substitution of links and nodes. In general it is not easy to calculate this distance and several algorithms have been proposed to implemented. See for instance~\citet{gao} for a survey.
However, in our setup, since the number of nodes $n$ is fixed and nodes are labeled, it becomes much easier. In particular, the edit operations are reduced to insertion or deletion of a link, and no isomorphisms are needed. 

Given two networks, $G$ and $H$ an edit path $\mathcal T = T_{1} T_{2} \cdots T_{k}$ between them is a sequence of edit operations that takes $G$ into $H$. Let $\Gamma(G,H)$ the set of edit paths between $G$ and $H$.
A cost is given to each edit operation, $c(T_i)$, representing whether or not the edit operation is a strong modification of the network. Then, the edit distance is defined as
$$
d(G,H)= \min_{\mathcal T \in \Gamma(G,H)} \sum_{i=1}^k c(T_i).
$$
If the cost of adding or deleting a given link is the same, the edit operations are flips $T_{ij}$ of the link $(i,j)$, which exchanges $1$ with $0$ on the $(i,j)$ entry of the adjacency matrix of the network, and the costs can be assigned to the links. In this case
\begin{equation}{\label{editpesada}}
d(G,H) = \sum_{i<j}\vert A_{ij} - B_{ij}\vert\, c_{ij}= \sum_{A_{ij} \neq B_{ij}} c_{ij},
\end{equation}
where $c_{ij}$ is the cost of adding or deleting the link $(i,j)$, and $A, B$ are the adjacency matrices of the networks $G$ and $H$ respectively. In particular, if all the costs are equal to $1$, then it is just the $L^1$ distance between the adjacency matrices $A$ and $B$. In order for $d$ to be a distance, we assume that all costs $c_{ij} > 0$. In what follows, the space of networks with $n$ nodes endowed with the metric \eqref{editpesada} is denoted by $\mathcal G$.

Another interesting distance, but computationally very expensive,  have been recently introduced by~\citet{pignolet} that takes into account the order in which the edit operations are executed. More precisely, the distance is defined as follows. Consider the subset of pairs of networks $ \mathcal M \subset \mathcal G \times \mathcal G$ of the form $(G, T(G))$, i.e., those which can be converted into each other with only one graph edit operation. To each pair $(G, T(G)) \in \mathcal M$ assign a cost, given by $c: \mathcal M \to \R^+$. Then, the distance is defined as
$$
d(G,H) = \min_{\mathcal T \in \Gamma(G,H)} \sum_{i=0}^{k-1} c(\mathcal T_i(G), \mathcal T_{i+1}(G)),
$$
where $\mathcal T_0(G)=G$ and $\mathcal T_i = T_1 \cdots T_i$. The inconvenience of this distance is that the number of costs that must be known is $\#  \mathcal M = m 2^{m}$. For this reason, it is important in this case to have an explicit formula for $c$, but in general it is not clear what would be adequate for real network data.

All the results that follow can be developed using any distance between networks. However, we work with the edit distance \eqref{editpesada} for which we obtain explicit algorithms of low computational cost (of order $n^2$) to develop the statistical problems addressed.

\subsection{Centers, and scale measure}
The notions of center and variability around it are fundamental to describe the distribution of a set of networks. Given that we have a metric in the space of networks there is a simple way to define them. In what follows we describe how to calculate this two important summarizing quantities.

We use boldface typeface for random elements. Let $p_G = \Pr{\G = G} $. The expected distance from a network $H$ to a random network $\G$ can be computed as
$$
\Ex{d(\G, H)} = \underset{G \in \cG} {\sum}  d(G,H)  p_{G}.
$$

\begin{definition}\normalfont
The \textit{central (or median) set} $\C$ of a random network $\G$ is the Frechet center with respect to the metric $d$, that is the subset of networks
\begin{equation}\label{esperanza}
\C: = \underset{H \in \cG}{\arg\min}  \ \Ex{d(\G, H)} .
\end{equation}

\end{definition}

The notion of median subset corresponds to minimizing the expected $L^1$--distance. A notion of mean subset can be defined by minimizing the expected $L^2$--distance. In the Euclidean setup they correspond to the $L^1$--median and to the usual expected value respectively. Given a sample $G_1, \ldots, G_\ell$ of random networks in $\cG$, applying definition \eqref{esperanza} to the empirical distribution, the notion of empirical center is obtained. More precisely,

\begin{definition}\normalfont
The \textit{empirical central set} $\hat{\C}_\ell$  is defined as the subset of networks fulfilling
\begin{equation}\label{mediana}
\hat{\C}_\ell  = \underset{H \in \cG}{\arg\min} \frac{1}{\ell}\overset{\ell}{\underset{i=1}{\sum}} d(G_i,H).
\end{equation}
\end{definition}

In general, the sets $\C$ and $\hat{\C}_\ell$ contain only one network, i.e., there exists a unique network that minimizes equations \eqref{esperanza} and \eqref{mediana} respectively. In this case, we call the unique central network the \textit{skeleton network} and we denote it by $S$ (or $\hat S_\ell$ in the empirical case). Note that in contrast to what is usually called the empirical average network (the sum of the adjacency matrices divided the number of networks) the skeleton belongs to the space of networks. 

The following proposition gives necessary and sufficient conditions for uniqueness of the central network, together with a complete characterization of the skeleton network and the subsets $\C,\hat\C_\ell$ when we have more than one solution. In the last case, the elements of the set form a lattice and there is a network in $\C$ (respectively in $\hat\C_\ell$) with the minimum number of links and another one with the maximum number of links. These are called the minimal and maximal centers respectively.

\begin{proposition}[Characterization of the central set]\label{centro}
\hfill\vspace{-0.1em}
\begin{enumerate}
\item[a)] $\C$ has a unique network if and only if $\,\Pr{\A_{ij} =1} \neq 1/2 \;\; \forall i,j$. The adjacency matrix of $S$ satisfies $A^{S}_{ij} = 1$ if and only if $\;\Pr{\A_{ij} =1} > 1/2$.
\item[b)] $\hat\C_\ell$ has a unique network if and only if $(1/\ell) \sum_{k=1}^\ell A_k(i,j)  \neq 1/2 \;\; \forall i,j$. The adjacency matrix of $\hat S_\ell$ satisfies $A^{S}_{ij} = 1$ if and only if $\;(1/\ell)  \sum_{k=1}^\ell A_k(i,j) > 1/2$.
\item[c)] Otherwise, for some pair $(i,j)$ we have $\Pr{\A_{ij} =1} = 1/2$. Define the minimal and maximal centers, $S$ and $L$, to be the networks whose adjacency matrices respectively satisfy that $A^{S}_{ij}=~1$ if and only if $\Pr{\A_{ij} = 1} > 1/2$, and $A^{L}_{ij} = 1$ if and only if $\Pr{\A_{ij} =1} \geq 1/2$. Then, the set $\C$ contains exactly all subnetworks of $L$ for which $S$ is a subnetwork. The same is true for the empirical version mutatis mutandis.
\end{enumerate}
\end{proposition}
The relevance of this last result is that it avoids finding the minimum by exhaustive search. If the empirical central network is unique then it contains only the links that are observed more than half of the time. Moreover, since the space of networks $\cG$ is finite, the law of large numbers follows immediately (see for instance \citet{breiman}, page 113 for the ergodic theorem for stationary sequences).

\begin{theorem} \label{consistencia}
Let $\G$ be a random network with law $\mu$ such that the central set $\C$ has only one element $S$. Let $\{\G_t, t \geq 1\}$ be a stationary and ergodic sequence of random networks with law $\mu$. If $\hat S_\ell$ is any element of the empirical central set $\hat \C_\ell$, then almost surely
$$\lim_{\ell \to \infty} d(S, \hat S_\ell) = 0.$$
In other words, the set of empiric central networks coincides with the skeleton network if $\ell$ is large enough. If the central set $\C$ has more than one element, we have that $\hat S_\ell \in \C$.
\end{theorem}

We are interested in a measure of the ``homogeneity'' (variability) of a random network. The most natural notion of dispersion associated with our problem is the following.

\begin{definition}\label{escala}\normalfont
Let $S^* \in \C$. The \textit{scale} of the random network $\G$ is defined as
$$
\sigma := \Ex{d(\G, S^*)} .
$$
\end{definition}
The corresponding empirical scale measure $\hat \sigma_\ell$ based on the sample $G_1, \ldots,G_\ell$, is obtained by replacing the expected distance by $(1/\ell)\sum_{i=1}^\ell d(G_i,S^*)$ in Definition \ref{escala}. We can derive strong consistency of $\hat \sigma_\ell$ to $\sigma$ from Theorem \ref{consistencia}, using the inequality
$$
\frac{1}{\ell}\sum_{i=1}^\ell \vert d(G_i, \hat S_\ell) - d(G_i, S) \vert \leq \max_{H \in \mathcal G} \vert d(H, \hat S_\ell) - d(H, S) \vert.
$$

We finish this section presenting two examples to illustrate the proposed framework.

\begin{example}\normalfont
An important distribution that arises in the space of networks $\cG$ is the (double) exponential type distribution given by 
\begin{equation} \label{expo}
\Pr{\G=H} = z e^{-\lambda d(H,S_0)}.
\end{equation}
If all costs are equal, the normalizing constant $z=e^{\lambda m}(1+e^{\lambda})^{-m}$, $\lambda >0$ is a parameter,  and $S_0$ is a particular network.  As in the real double exponential distribution, this law is symmetric, it has an explicit symmetry center and mode ($S_0$), and has an exponential decay. It is a particular case of the so called Exponential Random Graph Model~\citep{expon1,expon2}, and presents a unique central network. It is easy to show that it verifies $S=S_0$,  and $\sigma = m/(1+e^\lambda)$.

Note  that the empirical center given in equation \eqref{mediana} can be seen as a \textit{maximum likelihood} estimator of the center of the previous distribution. Indeed, if $G_1, \ldots G_\ell$ are i.i.d.\ random networks with this $\mu$ distribution, the empirical center coincides with the maximum likelihood estimate of $S_0$.
\end{example}

\begin{example}\normalfont
In the  Erd\H{o}s--R\'enyi model each link is present with a fixed probability $p$, independently.  Unlike the previous example, the well known Erd\H{o}s--R\'enyi model does not favor the presence of any link (nor a group a links), it is an homogeneous model. All networks with $k$ links are equally likely in the sense that they have the same probability.  If $p < 1/2$ the network is sparse. On the contrary, if $p > 1/2$ there exists a giant component, while $p = 1/2$ is the critical value.  Our notion of center captures the homogeneity and symmetry of the model, and provides a natural notion of center.  The central network is the empty network $G_{\emptyset}$ (the network with no links) if $p < 1/2$, in the case $p > 1/2$ it is the complete network $\G_\Omega$, and if $p = 1/2$ we get the entire space of networks.   

For this model, the scale is $\sigma= (1/2-|p-1/2|)m$ which has its maximum at $p = 1/2$, and it is zero for $p$ equal 0 and 1. This result is intuitive since the entropy is maximum at this value.
\end{example}
\begin{remark}\normalfont
A  different empirical notion of center has been introduced in \citet{ieee2001}, assuming that all costs are equal. It restricts the search of the minimizer to the networks in the sample, i.e., the center $\hat {R}_\ell$ is $\arg\min_j (1/\ell) \sum_{i=1}^\ell d(G_i,G_j)$.  The population version corresponds to minimizing the expected distance over the support of the underlying distribution $\mu$ of $\G$, that is the center is defined as $\arg\min_{H \in supp(\mu)} \ \Ex{d(\G, H)} $. If the support of $\mu$ is the whole space of networks $\cG$ both notions coincide. However, like in the case of high dimensional data, maximizing just over the sample is not a good strategy. Indeed, for example, it is easy to verify (using Hoeffding's inequality) that for the Erd\H{o}s--R\'enyi model with parameter $p< 1/2$,
\begin{align*}
\Pr{\hat S_\ell \neq G_{\emptyset}} &\leq 1 - \left(1-e^{-2\ell(p-1/2)^2}\right)^m, \\
\Pr{\hat{R}_\ell \neq G_{\emptyset}} &\geq \prod_{i=1}^\ell \Pr{\G_i \neq G_\emptyset} = (1 - (1-p)^m)^\ell.
\end{align*}
Thus, $\hat S_\ell$ converges at a much better rate.
\end{remark}

\section{Depth function}
%==============================
In this section we first introduce a notion of depth in the space of networks. A depth function is a function that orders the space in terms of center-outward position. The idea has been introduced in the robust statistics literature. The most well known depth notions for the Euclidean space are the half--space depth \citep{tukey}, simplicial depth \citep{liu}, the $L^1$--depth \citep{brown,small}, the projection depth \citep{arcones2006},  and Mahalanobis depth \citep{mahala}. Several important applications to different statistical problems based on depth concepts have been developed in the last years.

Given a fixed network $H$ and a sample of random networks $G_1, \ldots, G_\ell$ with the same distribution we consider the $L^1$--depth notion with respect to the metric $d$, which in particular defines the central network (also called spatial median) in our setup. The central set corresponds with the set where this depth is maximized. More precisely,

\begin{definition}\normalfont
We define the empirical depth at the network $H \in \cG$, as
$$\hat{D}_\ell(H) =  m - \frac{1}{\ell} \overset{\ell}{\underset{i=1}{\sum}} d(G_i, H),$$
which corresponds to the population depth given by
$D(H) = m - \Ex{d(\G, H)} $.
\end{definition}

Observe that both the empirical and population depth are non--negative, and fulfill the main properties of a depth function given in \citet{serfling}. Moreover, we have a simple explicit solution for the median center maximizing $D(H)$ given by Proposition \ref{centro}. On the contrary in the Euclidean space,  an optimization method is required to maximize the $L^1$--depth. In fact, a fast monotonically convergent algorithm to calculate the $L^1$--median of a data set in $\R^d$ has been proposed in \citet{vardi}.

An important property of Definition 4 is that the depth function determines the network law. This result, that follows from the invertibility of distance matrices given by \citet{auer}, has an important impact in statistics and in particular in our setup since it allows to develop statistical methods based on the depth $D$ and obtain results for the space of networks $\cG$. 
\begin{proposition}\label{depth_prob}
Given two distributions $\mu,\nu$ on $\cG$. Write $D_\mu(H) = m - E_\mu(d(\G, H))$ to explicitly note the dependency. Then, $\mu = \nu$ if and only if $D_\mu(H) = D_\nu(H)$ for all $H \in \cG$.
\end{proposition}

In general, depths do not determine distribution, the only known result is for Tukey's half space depth when the measure is discrete and can be found in \citet{cuesta}. The empirical depth function $ \hat{D}(H) $ converges almost surely to the population version $D(H)$ uniformly. Indeed, as a consequence of the ergodic theorem we have the following theorem.

\begin{theorem}[Uniform convergence of the depth function]\label{consistenciadepth}
Given a stationary ergodic sequence of random networks $\{\G_t: t \geq 1 \}$ with common law $\mu$. Then, almost surely as $\ell \to \infty$, 
$\max_{H \in \cG} \vert \hat{D}_\ell(H) -D_\mu(H)\vert \to 0$.
\end{theorem}

Moreover, the asymptotic distribution of the depth process is Gaussian.

\begin{theorem}[Asymptotic normality of the depth process]\label{normalidad}
Given a strictly stationary $\alpha$--mixing sequence of networks $\{ \G_t: t \geq 1\}$ in $\cG$ with common distribution $\mu$ fulfilling $\sum_{n=1}^\infty \alpha(n) < \infty$. Fix an ordering $(G_j)_{j=1,\dots,2^m}$ of the elements of the space $\cG$ and define 
$$
\Z_\ell = \left( \hat{D}_\ell(G_j) - D_{\mu}(G_j) \right)_{j=1,\dots,2^m}, \quad
\Y_k = \left(d(\G_k, G_j) - \Ex{d(\G_k,G_j)}\big. \right)_{j=1,\dots,2^m}.
$$
If in addition
$\sum_{k=1}^\infty \beta_0^T \Ex{\Y_1^T\Y_k} \beta_0 > 0$, for all $\beta_0$ with $\Vert \beta_0 \Vert = 1,$, then, $\sqrt{\ell} \beta_0^T \Z_\ell$ converges weakly as $\ell \to \infty$ to a normal distribution with mean zero and with the same variance as $\beta_0^T \Y_1$.
\end{theorem}

\section{Hypothesis Testing: Test based on Random Projections.}
In this section we study the one and two samples hypothesis testing problem. Here we are interested in comparing for example two samples of networks and determine if they belong to the same population. A naive approach is to test directly the two distributions (or their corresponding depths, see Proposition~\ref{depth_prob}) with a Chi-square or a Kolmogorov type test based on the statistic $\max_{H \in \cG} \vert \hat{D}_\ell(H) -D_\mu(H)\vert$. However, the computational cost is of order $2^m$ and it requires a huge sample size in order to have a large power.
On the other side, tests based on random projections for the one and two samples problems, as those considered in \citet{ric2,ric1},
 can be implemented in our setup since the space of networks $\cG = \{0, 1\}^m$ is contained in $\R^m$. They are based on the use the following Corollary which we describe briefly. Let $\mu$ be a Borel probability measure on $\R^m$ ($m \geq 2$). Given a direction $h \in \R^m$ , let $\langle h \rangle$ be the one--dimensional subspace spanned by $h$ and $\mu_{\langle h \rangle}$ the distribution of the orthogonal projection of $\R^m$ onto $\langle h \rangle$, i.e.,
$\mu_{\langle h \rangle}(B) = \mu(\pi^{-1}_{\langle h \rangle} (B))$, for Borel $B \subset \R$.

\begin{proposition} [\citealp{ric2}, Corollary 3.2]   Let $\mu, \nu$ be Borel probability measures on $\R^m$, where $m \geq 2$. Then $\mu = \nu$ if
\begin{itemize}
\item[a)] the absolute moments $\Theta_n = \int \| x \|^nx\, d\mu(x)$ are finite and satisfy the Carleman condition $\sum_{n\geq 1} \Theta_n^{-1/n} = \infty$,
\item[b)] the set $\mathcal{E}(\mu,\nu)=\{h \in \R^m: \mu_{\langle h \rangle} = \nu_{\langle h \rangle}\}$, has positive Lebesgue measure.
\end{itemize}
\end{proposition}
It is clear that any distribution in $\R^m$ supported on $\{0, 1\}^m$ fulfills Carleman condition and the previous result holds. For statistical purposes this means that in order to address the one sample problem $H_0: \mu=\mu_0$ vs $H_A: \mu\neq \mu_0$, or the two samples problem $H_0: \mu=\nu$ vs $H_A: \mu\neq \nu$, where $\mu$, $\mu_0$, and $\nu$ are distributions in the space of networks, a standard one dimensional test is enough. For example, a Kolmogorov--Smirnov test based on the projected data onto one randomly chosen (for instance uniformly on the unit sphere) is able to test the null hypothesis presented above. 

To improve the power, we may use many random projections onto i.i.d.\ uniformly distributed directions $h_1,...,h_q$. 
For the two samples the procedure is as follows  (the one sample problem is completely analogous). 

Given two independent iid sequences of networks $\mathcal{G}_{1} = \{\G^1_1,...,\G^1_{\ell_1}\}$, $\mathcal{G}_{2}= \{\G^2_{1},...,\G^2_{\ell_2}\}$ and $q$ random directions $\{\mathbf{h}_1,...,\mathbf{h}_q\}$ on $\R^m$, define for each $i=1,\dots,q$
$$
\mathbf{X}_{ij} =  \langle \mathbf{h}_i,\G^1_{j} \rangle,  \quad  j=1,\dots, \ell_1;\qquad
\mathbf{Y}_{ij} =  \langle \mathbf{h}_i,\G^2_{j} \rangle,  \quad j=1,\dots, \ell_2.
$$
For each $i$ consider the samples $\{\mathbf{X}_{ij} \ \ j = 1,...,\ell_1\}$, $\{\mathbf{Y}_{ij}  \ \ j = 1,...,\ell_2\}$ and define the binary
random variable $\mathbf{Z}_i$ taking value one if the two samples Kolmogorov--Smirnov test of level $\alpha$ reject the null hypothesis and zero otherwise.
Let $\mathbf{Z} = \sum_{i=1}^q \mathbf{Z}_i$. Under the null assumption, $\mathbf{Z}$ has a Binomial($q,\alpha$). Therefore, the exact $p$--value= $\Pr{\mathbf{Z} > a}$ can be computed. If instead of two independent iid sequences of networks we have two independent stationary and ergodic sequences, the Glivenko--Cantelli Theorem still holds (see for instance Theorem 1.1 on page 4, in \citep{MS}), and we can use the  bootstrap results for stationary and ergodic sequences given in \citet{ahmetetal} 
to approximate the critical value of the Kolmogorov--Smirnov test of level $\alpha$, or for stronger results assuming mixing conditions or even weaker dependence conditions, see for instance \citet{dehling} and \citet{doukhan} respectively for the bootstrap validity.

\section{Sparse Principal Components}
Principal components is an important statistical tool when analyzing data, particularly for high dimensional and functional data~\citep{greco}.  The objective of this technique is to reduce the dimension $p$ of the data using linear combinations of the variables. This is done by projecting the data onto  the $k\ll p$ dimensional subspace which minimizes the distance to the original random vector. Equivalently, the principal components can be defined iteratively. The first is the direction on which the projection of the random element has maximal variance.  The next one, maximizes the variance of the projection on the orthogonal subspace to the first one and so on. The absence of projections in metric spaces makes the extension non trivial. Since the space of networks is very large we want a sparse notion of principal components. In what follows we introduce some methods in such direction for random elements in the space $\cG$.

Let $G_\emptyset$ be the empty network and write $|H| = d(H,G_\emptyset)$ for the sum of the costs of links in $H$. Given $G,H\in\cG$ define the intersection network $G \wedge  H$ as the network with only the common links to both. Note that $|G\wedge H|$ is the weighted inner product, given by the costs, between the adjacency matrices of $G$ and $H$. Recall that, given $a,b,x$ arbitrary points in a metric space, we say that $x$ belongs to a geodesic from $a$ to $b$ if $d(a,b)=d(a,x)+d(x,b)$. So, if $d(G,H)=k$ then there are $k!$ geodesics between them. Given a network $H$ we define the set $\cS(H)$ of all geodesics curves in the space $\cG$ joining $H$ with the complete network, denoted by $G_{\Omega}$. In other words, $\cS(H)$ is the set of all networks which have $H$ as a subnetwork.

Given a random network $\G$, we define the first \textit{principal component} as the set of networks $\Q_1$ that maximize the variance of the following ``projection"
\begin{equation} \label{pp1}
\Q_1=\underset{Q \in \cG} {\arg\max} \ \Var\left( \frac{|\G\wedge Q|}{|Q|} \right).
\end{equation}
If $\mathcal Q_1= \{Q_1, \ldots, Q_p\}$ then the \textit{principal component space} is $\cS_1 = \cup_{i=1}^p \cS(Q_i)$. Typically the set $\Q_1$ is a single network. The analogue of having more than one network in $\Q_1$ in the classical Euclidean case is having eigenvalues of the covariance matrix with multiplicity greater than one.

To define the \textit{second principal component}  $\Q_2$  we consider the same problem, but now we maximize the variance within the ``orthogonal" subset $\cG \setminus \cS_1$.

Observe that $H \in \cG \setminus \cS_1$ iff $H$ has no links in common with any element in $\Q_1$, i.e., it has no link in common with the network $\tilde Q_1:= Q_1 \vee \ldots \vee Q_p$, which contains all links present in at least one $Q_j$, $j=1, \ldots p$. In this sense we refer to $\cG \setminus \cS_1$ as the orthogonal subset. This particular network, $\tilde Q_1$, can be considered as the most informative network to visualize $\cS_1$. The set $\cG\setminus \cS_1$ has cardinality $2^{m-|\tilde Q_1|}$. The next principal components are defined analogously.
To define the corresponding empirical version of expression \eqref{pp1} let
$$
\Delta_\ell(Q)= \frac{1}{\ell} \sum_{i=1}^\ell \frac{\left(|G_i \wedge Q|- \Lambda_\ell(Q)\big.\right)^2}{|Q|^2},
$$
where $\Lambda_\ell(Q)=(1/\ell) \sum_{k=1}^\ell |G_k \wedge Q|$. The empirical first principal component is the set of networks that maximizes the empirical variance ($\Delta_\ell(Q)$), and the empirical first principal component space is the set of networks that have the links of the first principal networks, i.e. $\hat{\cS}_1= \cup_{i=1}^p \cS(Q_i)$ where $Q_i$ verifies $\Delta_\ell(Q_i)$ is maximum.

%%%%%%%%%%%%%%%%%%%%%%%%%%%%%%%%%%%%%%%%%%%%%%%%%%%%%%%%%%%
\begin{proposition}[Characterization of the sparse principal component sets]\label{compocompleto}
Let $E_{ij}$ be the network consisting of only the $(i,j)$ link and $p_{ij} = \Pr{\A_{ij}=1} $. If
$$
M_1=\{(i,j): 1\leq i< j \leq n\} 
\quad\text{and}\quad 
L_1 = \underset{(i,j) \in M_1}{\arg \min}\; |p_{ij}-1/2|,
$$
then $\cS_1 = \cup_{(i,j)\in L_1} \cS(E_{ij})$. Analogously, if
$$
M_k=\{(i,j):\; A^{H}_{ij}=0 \;\;\forall H \in \cup_{i=1}^k \cS_i \} 
\quad\text{and}\quad 
L_k = \underset{(i,j) \in M_k} {\arg \min}\; |p_{ij}-1/2|,
$$ 
then $\cS_k = \cup_{(i,j)\in L_k} \cS(E_{ij})$.
\end{proposition}

Based on the previous proposition, a simple algorithm of order $n^2$ is presented below.\\

%\noindent\fbox{\parbox[b][8.8em][t]{\textwidth}{		
%\vspace{-0.1em}
\begin{itemize}[leftmargin=4mm]
\item \textbf{Step 1:} Compute $W(Q_i):=\vert \Var (|\G\wedge Q_i|/|Q_i|)  - 1/2 \vert$ for all $Q_i$ with a single link.
						
\item \textbf{Step 2:} Let  $W(Q_{(1)}) \geq \ldots \geq W(Q_{(M)})$ be the order statistics.
						
\item \textbf{Step 3:}   If there are no ties, the first principal space is the set of networks with $Q_{(1)}$ as a subnetwork, the second one is the set of networks for which $Q_{(2)}$ is a subnetwork and $Q_{(1)}$ is not, and so on. If there are ties, say for instance $W(Q_{(j_1)})= W(Q_{(j_p)})$, the principal space is just the union of the principal spaces of each $Q_{(j_r)}$, for $r=1, \dots, p.$			
\end{itemize} %}}
\smallskip

%%%%%%%%%%%%%%%%%%%%%%%%

\begin{remark} Our definition of sparse principal components normalize the data by $\vert Q \vert$. The same normalization is used when dealing with directional data, and we believe it is adequate in our context.
Indeed, if in definition \eqref{pp1} we normalize by $\sqrt{\vert Q\vert}$ instead of $\vert Q \vert$, that is
\begin{equation}\label{pp2}
\Q_1=\underset{Q \in \cG} {\arg\max} \ \Var\left( \frac{|\G\wedge Q|}{\sqrt{|Q|}} \right),
\end{equation}
we obtain a less sparse version of the principal components. 
\end{remark}

\begin{proposition}[Consistency of Principal Components]\label{compoconsist}
Given a stationary and ergodic sequence of random networks $\{\G_t, t \geq 1\}$, the empirical principal components as well as the principal components sets converge a.s. to their corresponding population versions.
\end{proposition}

\begin{example}\normalfont
Distributions with spherical symmetry satisfy do not have a principal direction, since all directions are equally informative. The distribution given in equation \eqref{expo} is one such example. However, a mixture of two or more distributions of that form breaks the symmetry.
Here we consider a mixture of 3 exponentials with the same $\lambda$, whose centers are shown in Figure~\ref{compo} (A), to generate a sample of size 1000. We consider equal costs for each link. The measure is
\begin{equation}\label{mix}
\mu(H) = \Pr{\G = H}  =\overset{3}{\underset{i=1}{\sum}}p_ic e^{-\lambda d(H,S_i)},
\end{equation}
where $(p_1, p_2, p_3)=(0.4,0.3,0.3)$.

\begin{figure}[h]
\centering
\hspace*{-0.5cm}
\includegraphics[angle=0,width=1.0\textwidth]{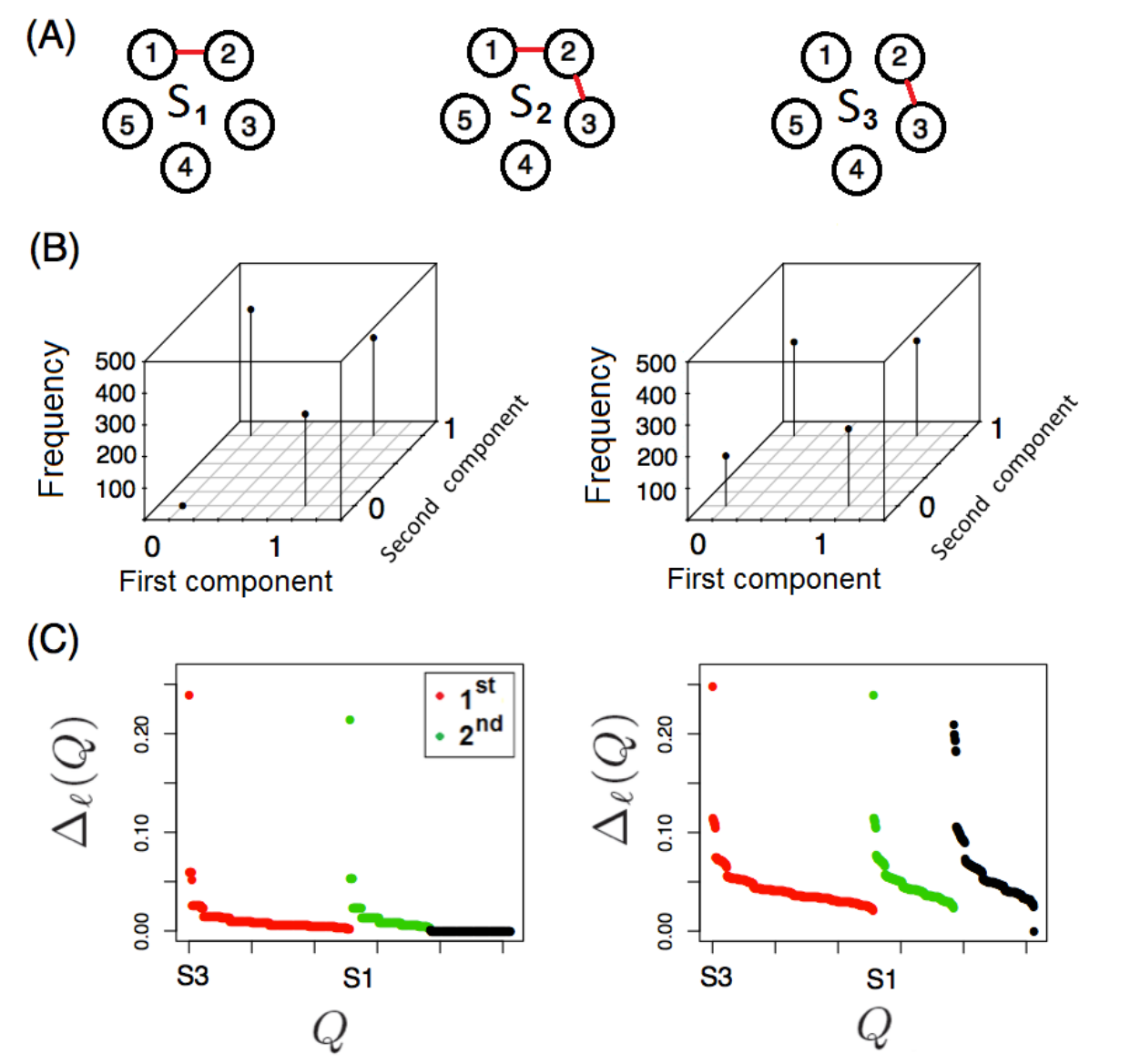}
\caption{(A) Networks used in the definition of $\mu$. (B) Projected data on the first ($S_3$) and second principal components ($S_1$)  are shown for a one thousand sample generated using equation~\ref{mix} with $\lambda$ equal 10 (left panel), and 1 (right panel). (C) Empirical variance, $\Delta_\ell(Q)$, of the sample of networks projected on each network $Q$.   Red (Green) points represent all the networks that belong to the first (second) principal component space $\cS(S_3)$ ($\cS(S_1)$). On the left we show the results for  $\lambda$ equal 10 (left panel), and 1 (right panel). } 
\label{compo}
 \end{figure}
 In this model it is easy to see that 
%\begin{equation} 
\begin{align*} 
%\begin{split} 
% \underset{\lambda \rightarrow \infty}{\lim} \mu(S_1)=0.4, \underset{\lambda \rightarrow \infty}{\lim} \mu(S_2)=0.3, 
%\underset{\lambda \rightarrow \infty}{\lim} \mu(S_3)=0.3, \\ 
 \lim_{\lambda \rightarrow \infty} \Pr{\A_{12} =1} =0.7, \lim_{\lambda \rightarrow \infty} \Pr{\A_{13} =1} =0.6, \lim_{\lambda \rightarrow \infty} \Pr{\A_{ij} =1} =0 \quad \mbox{with} \quad ij \notin\{13,12\}.
\end{align*}
%\end{equation}
%\end{split}  
 Also it easy to see that for fixed $\lambda$ these probabilities verify$$
 |\Pr{\A_{13} =1}- 0.5|<|\Pr{\A_{12} =1}-0.5| < |\Pr{\A_{ij} =1}-0.5| \quad \mbox{for} \quad ij\notin\{13,12\}.
 $$
Therefore the first principal component space is generated by the one link network $S_3$, and the second principal component space is generated by the $S_1$ network. For $\lambda=10$ these probabilities become $\Pr{\A_{12} =1}= 0.69998 $, $\Pr{\A_{13} =1}=0.5999909$, $\Pr{\A_{ij} =1}=0.0000454 \ \mbox{for} \ ij\notin\{13,12\}$, and for $\lambda=1$ the values are 0.5924234, 0.5462117, and 0.2689414 respectively.
 
 Figure~\ref{compo} (B)  shows the projection of a network sample of size one thousand generated from the exponential model presented in equation \eqref{mix} with $\lambda$ equal 10 (left panel) and 1 (right panel). For both $\lambda$ values we compute the dot products $|G_i \wedge S_3|$ and $|G_i\wedge S_1|$ for $i=1,\dots,1000$. These dot products only take the values 0 or 1 depending on whether the network contains the corresponding link or not. The height in the $z$-axis shows the number of networks (frequency) in the sample projected at each of the four possibilities. Note that if we just observe the data in the first component the frequencies of ones and zeros are closer to each other than when the data is observed in the second component. Similar frequency values for ones and zeros corresponds to a greater variance. 
Both empirical components coincide with the population version. For $\lambda=10$, the proportion of networks in the sample containing the link $(1,3)$ is $0.598$ while for $(2,3)$ is $0.706$. For $\lambda=1$, the proportion of networks in the sample containing the link $(1,3)$ is $0.54$ while for $(2,3)$ is $0.574$. These proportions are very similar to their theoretical values. 
Figure~\ref{compo} (C)  shows the empirical variance, $\Delta_\ell(Q)$, of the projected sample of networks over each network $Q$.  The networks are represented in the $x$-axis and they are sorted by the appearance of the components. The first component is the network $S_3$ and this one generates all the first principal space $\cS(S_3)$ (red points).  The second component is $S_1$ and it generates all the second principal space $\cS(S_1)$ (green points). The figure shows the results for $\lambda$ equal 10 (left panel), and 1 (right panel). Note that as $\lambda$ goes to zero the first, second and the subsequent components degenerate. Therefore in the empirical case, we have to deal with the problem of identifying if the differences observed, for example between $\Delta_\ell(S_1)$ and $\Delta_\ell(S_3)$, are a product of randomness or the are really different. For a sample size $\ell$, it is possible to define $\eps_\ell$ such that the first principal component networks are those that verify 
$$\hat{\Q}_{1}=\{Q \in \cG: \Delta_\ell^{\max} - \Delta_\ell(Q) <\eps_\ell \},$$
where $\Delta_\ell^{\max}=\max_{Q\in \cG} \Delta_\ell(Q)$.
\end{example}

\section{Other multivariate techniques based on distances}
%==============================
In this section we briefly discuss some other important multivariate methods. It is clear that extending to our setup other techniques that only rely on distances such as clustering, classification, multidimensional scaling and nonparametric regression is mainly straightforward. The choice of the distance $d$ for these methods is very important.  In applications we may be interested in taking into account some ``features'' of the network to improve the performance of the  procedure. A pseudo-metric $d_f$ typically is considered to capture closeness with respect to the selected ``features''. These features can be global, such as the diameter or average degree, or can be local, for example the degree of each node. A simple solution to take into account features is to work with the distance
$$\tilde d=\alpha d+ (1-\alpha) d_f\quad \text{(for $0<\alpha<1$)},$$  
which is still a metric. Asymptotic results for this distance still hold, while this is not the case if we just use a pseudometric. To illustrate we emphasize two main problems: clustering and classification of networks. The classical k-means algorithm for clustering, as well as k-nearest neighbor, for supervised classification can be applied directly.

\subsubsection{Unsupervised classification}
%===============
Let us suppose that the probability measure $\mu$ is such that $k$ groups or clusters of networks can be identified. Two networks in the same group are close together (similar), while networks that belong to different groups are far apart. We want to identify each of the $k$ groups. The most well known clustering methods are $k$--means or $k$--medioids, which are only based on the distances between the random elements. More precisely, the algorithm looks for the centers of the groups and then assign each data to its nearest center.

In our setting, what is important is to identify in a good way each of the $k$ center networks, that we denote by $S^*_1, S^*_2, \ldots, S^*_k$. The strategy proposed here is the same to that of k-means (k-medioids in our case). We look the $k$ networks that maximize the depth of order $k$ defined as:
$$
D_k(H_1, \ldots, H_k) = m - \Ex{ \min_{i=1,\dots,k}  d(H_i,\G)} ,
$$
i.e., we look for subsets $\{S_{1}^*, \ldots, S_{k}^*\}$ that satisfy
$$
D_k(S_{1}^*, \ldots, S_{k}^*)=  \underset{H_1,\ldots, H_k}{\max} \ D_k(H_1, \ldots, H_k).
$$
Then, each network is assigned to its nearest center and we obtain a partition of the space. The asymptotic results for $k$--means and $k$--medioids given in \citet{pollard} are valid for compact metric spaces, which covers our setup. In the empirical case we look for the empirical center networks $\hat{S}^*_1,\hat{S}^*_2,  \ldots, \hat{S}^*_k$ that maximize the empirical depth of order $k$,
$$
\hat D_k (H_1, \ldots, H_k) = m -  \frac{1}{\ell} \sum_{j=1}^\ell \min_{i=1,\dots,k}  d(H_i, G_j).
$$

\subsubsection{Supervised classification}
%===============
In this case we have a training sample $(\Y_1,\G_1), \ldots, (\Y_\ell,\G_\ell)$ where $\{\G_t: t\geq 1\}$ is a sequence of random networks and $\{\Y_t: t \geq 1\}$ stands for the labels that indicates to which subpopulation (group) the individual belongs. For binary classification $\Y_t \in \{0,1\}$ indicating sick or healthy for instance.
The problem consist on predicting the label of a new observation only based on $\G_{\ell+1}$ and the training sample.

The most simple and well known nonparametric classification method is $k$--nearest neighbors. The method just looks for the $k$ nearest neighbors of $\G_{\ell+1}$ among the sample $\{\G_t: 1 \leq t \leq \ell \}$ and assigns the label by majority vote within the labels of the $k$--nearest neighbors. It can be applied in our setup as it is just based on distances. The method is asymptotically optimal as long as $k=k(\ell) \to \infty$ and $k/\ell \to 0$ as $\ell \to \infty$ \cite[see for instance][]{devroye}.

\section{Case Studies}
%==============================
In this section we apply the results from the previous sections to data from two real networks. We present examples in two categories: networks of level 1 (L1 networks), which corresponds to the case where the data are directly a sequence of networks in time, and networks of level 2 (L2 networks), which are build up from a stochastic process given at each node. For instance, a link between two nodes is present if the correlation between the corresponding stochastic processes of the two nodes is greater than a given threshold. In this case, we lose information while trying to get a simpler model that still retains the relevant characteristics. Level 2 networks are a common tool in functional Magnetic Resonance Imaging (fMRI) studies where they are called functional networks.

\subsection{A L1 network example: Social Network}
%==============================
How a disease or information propagates in a population is a very relevant question. For many diseases face-to-face interactions are decisive in the propagation process. \citet{barrat} collected very interesting data to study face-to-face social interaction. By using wearable sensors they were able to describe how do French high school students socially interact \cite[see][for details]{barrat}. Each of the 327 students involved in the study had a sensor. The sensors exchange packets only when within 1--1.5 meters of one another, thus allowing to know which students are socially interacting in time. The data is publicly available at {\small\url{http://www.sociopatterns.org/datasets/high-school-contact-and-friendship-networks/}}, and corresponds to the face-to-face interactions (links between students) every 20 seconds during five days. For the application shown here we analyze only the last day studied by the authors (Friday) during all the school day. In the analysis we divide the day in four equal periods, which we label as Morning 1, Morning 2, Afternoon 1 and Afternoon 2.  We do not consider the times of ``arrival'' (8:00--8:30) and ``departure'' (16:30--17:00) from school, and we assume all costs are equal.

\begin{figure}[ht]
\centering
\includegraphics[angle=0,width=0.8\textwidth]{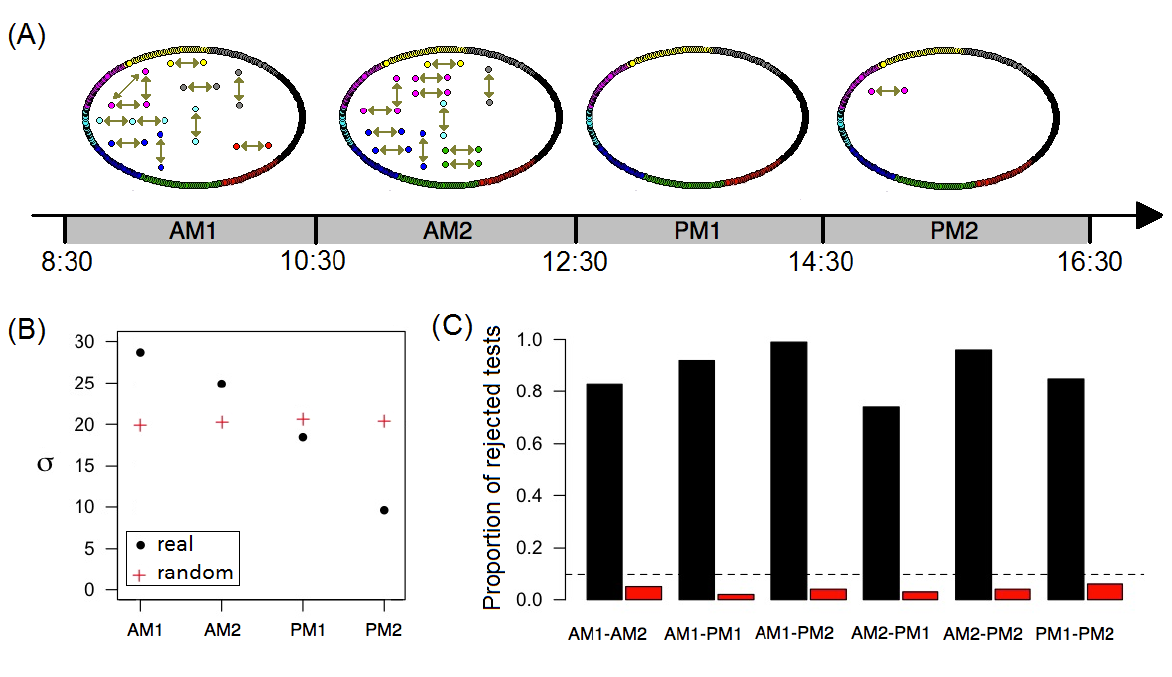}
\caption{(A) Empirical Central Graph and (B) Variability for each of the four time intervals AM1, AM2, PM1, and PM2. (C) Proportion of rejected tests, at a significance level of  0.05. The results for a random network are shown in red. }
\label{social}
 \end{figure}

For each time interval of 20 seconds we have a network, obtaining a sequence of 360 networks $G_1, G_2, ...,G_{360}$. We compute the empirical Central Graph (Figure~\ref{social} (A)) and the Graph Variability $\sigma$ (Figure ~\ref{social} (B)) for each sequence. The central networks for the periods AM1, AM2, PM1 and PM2 have few links, 12, 11, 0 and 1 respectively. Some of the links that join two students are present in both AM1 and AM2 periods (vertical links). Looking at  the central networks it seems that the students interact in a different way between the morning and the afternoon.

More precisely, the network variability decreases with time as it is shown in Panel (B). To understand how different are the network probability laws between the time intervals we test it by the random projection procedure described above. One hundred tests were performed, and the proportion of rejected tests (at the level of significance 0.05) is shown in Panel (C) for each pair of periods. In all cases, the proportion of rejected tests is much greater than the critical value (dashed line). The test is able to detect differences even in scenarios where there are few links in comparison with its maximum number (in this case 53301). Namely, the face-to-face interactions are different and this is detected because the law changes along the day.

Next, we compute the central networks, the network variability and the corresponding tests for a sequence of networks generated artificially. More precisely, a link is randomly assigned to one the 1440 time steps, and the same time categories are analyzed.  The central network is the null network in the four intervals, and the variability (red crosses in Panel (B)) takes a constant value approximately equal to 20. It is easy to show that this value is the one expected for the Erd\H{o}s-Renyi model, $\sigma= (1/2-|p-1/2|)53301$, for $p$ given by the average density of links in a time step (21.58/53301).  Finally,  in Panel (C) we show the proportion of rejected tests with red bars (which is clearly below the significance level) verifying that these randomly ``shuffled'' networks have the same distribution, as it is expected.

\subsection{A L2 network example: Correlation Network}
%==============================
Network techniques based on statistical associations between climate parameters at different points on Earth have been used recently \cite[see for example][]{donges,tsonis}. Networks are also used to characterize climate dynamics, and phenomena such as the El Ni\~no/Southern Oscillation \citep{gozolchiani}. As far as we know, most results in this area use statistical data to generate a unique network rather than analyzing a \textit{sample} of networks. Here we analyze climate data with the methodology developed in Sections 2-5.

\begin{figure}[ht]
\centering
\includegraphics[angle=0,width=\textwidth]{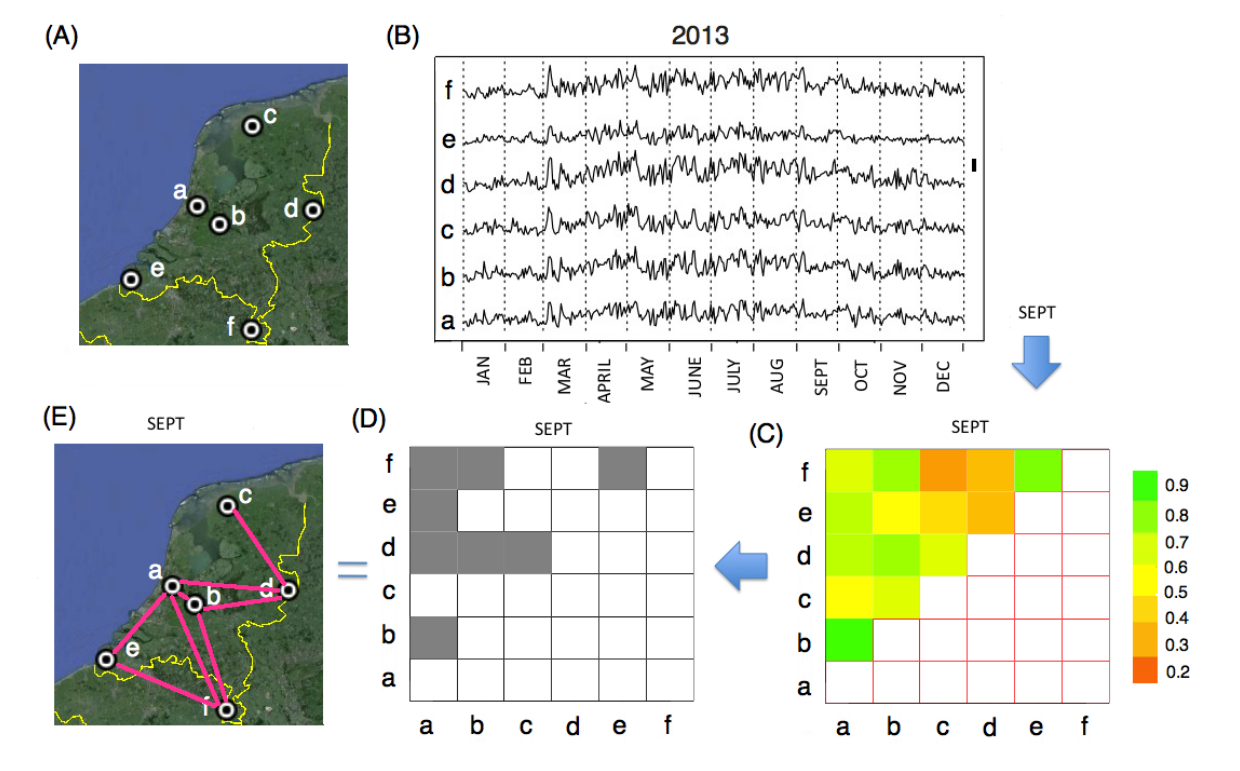}\vspace{-1em}
\caption{ (A) Location of the meteorological stations. (B) Thermal amplitude series for the year 2013. The segment on the right end represents 10 degrees Fahrenheit. (C) Correlations between the thermal amplitudes for September 2013. (D) Adjacency matrix obtained using a threshold of 0.48. (E) Associated network.}\label{stations}
\end{figure}

 The data consist on daily thermal amplitude values from six Dutch meteorological stations located across the country (North, North-West, North-East, Middle, West, South-West, South-East).  Figure~\ref{stations} (A)
 shows the location of each station studied. Figure~\ref{stations} (B) shows the daily thermal amplitude (temperature range) for each station during the year 2013.  For each month we compute the (Spearman) rank correlation matrix between the series, as shown in Figure~\ref{stations}~(C) (by symmetry only the upper values are depicted). We consider rank correlations in order to avoid sensitivity to a few days with extreme values.
Finally, a network is obtained by keeping the links with statistically significant correlations, corrected by multiple comparisons ($p$--values using a t-test for correlations $< 0.05/15$). Two meteorological stations are connected if they share similar thermal behavior. Figure~\ref{stations} (D) and (E) show the network obtained for September 2013. This method is a standard technique for constructing correlation or functional networks mostly used in Neuroimaging \citep{sporns} and Finance \citep{vespignani}.

\begin{figure}[h]
\centering
\includegraphics[angle=0,width=\textwidth]{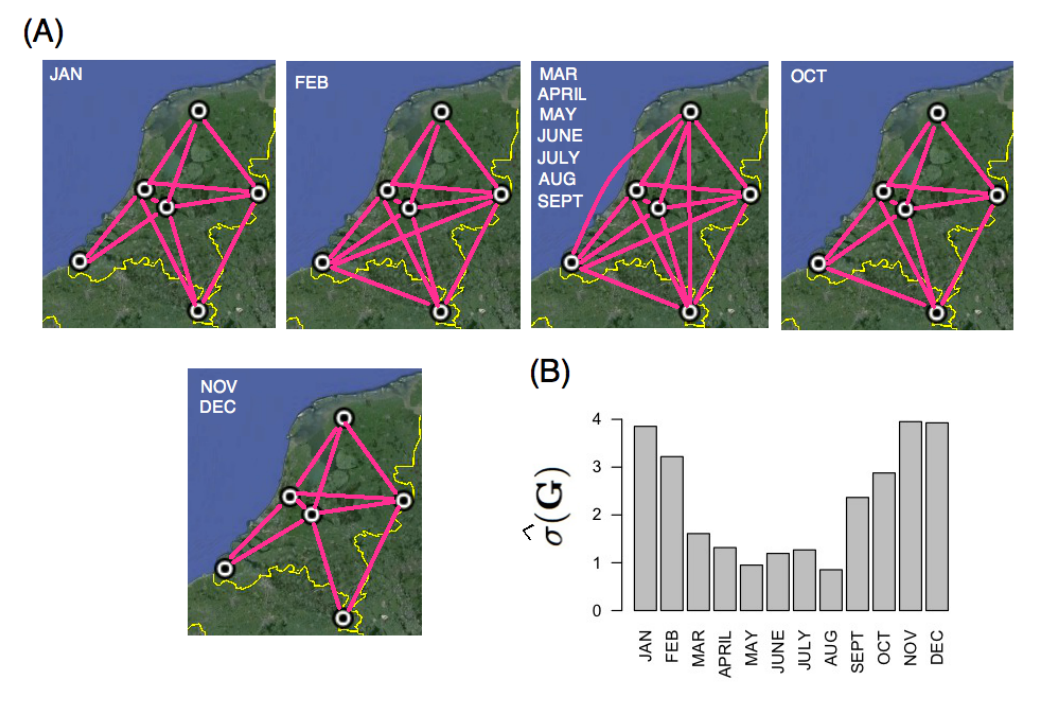}
\caption{(A) Central networks. (B) Scale values.}\label{holanda}
\end{figure}

Using the method described above we construct a network for each month from 1973 to 2013. The data can be obtained at {\small\url{http://www.ncdc.noaa.gov}}. We study the evolution over the last 41 years. Figure~\ref{holanda} (A) shows the central network for each month. The months of March, April, May, June, July, August and September share the same central network which is the complete network. Some of the links connecting geographically distant stations disappear during the coldest months. Moreover, each year the networks observed at each month are different. Figure~\ref{holanda} (B) shows the scale or variability for each month. Low temperature months present networks with more variability. Climate networks for months with colder temperatures (October to February) have in average less links and are more varying. 

\begin{figure}[h]
\centering
\hspace*{-0.5cm}
\includegraphics[angle=0,width=0.8\textwidth]{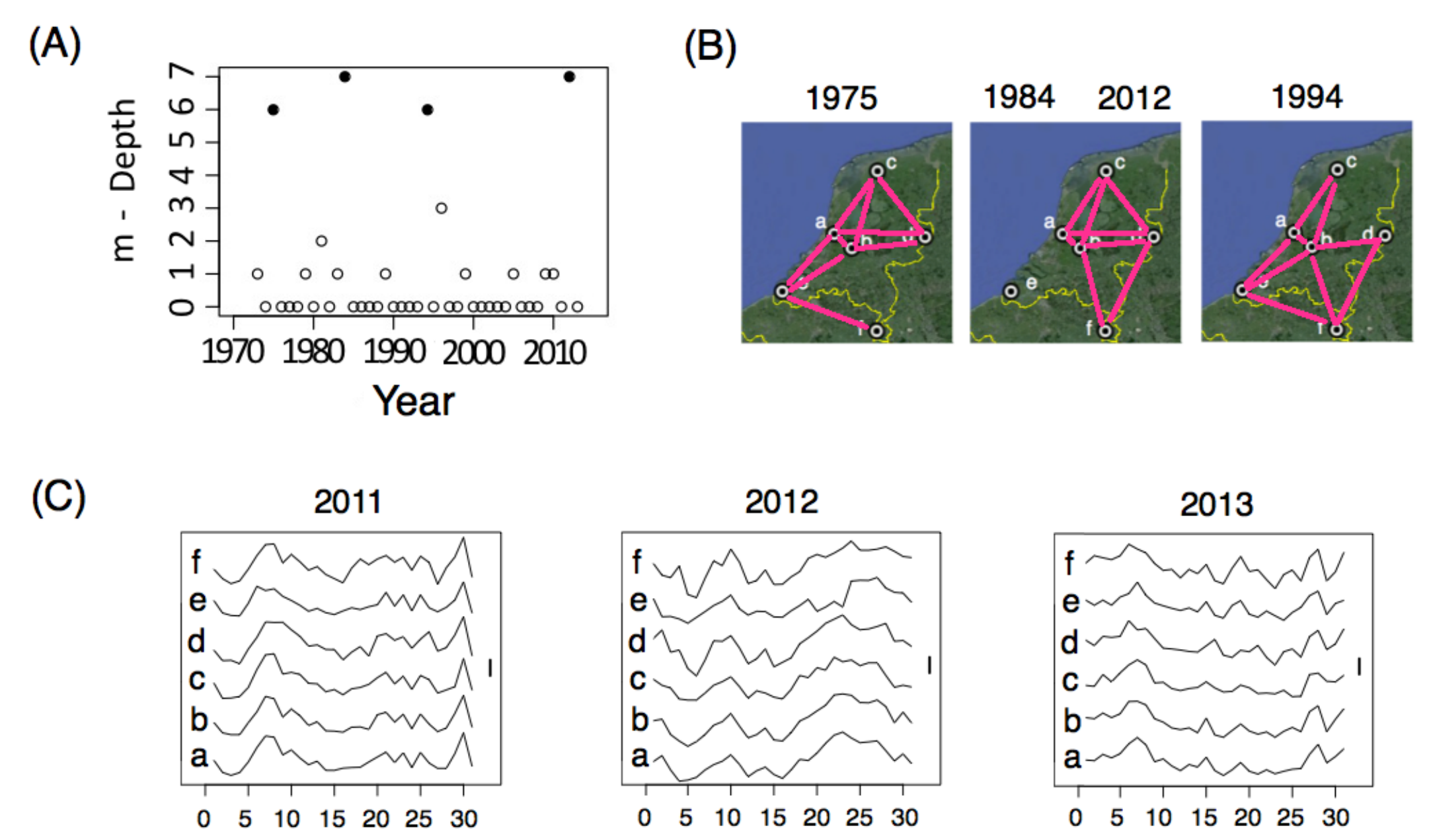}
\caption{(A) Empirical expected distance for each year at the month of May. Black points correspond to outliers.  (B) Atypical networks. (C) Thermal amplitude series for the six stations. The segment represents 10 degrees Fahrenheit.} \label{holanda_time}
\end{figure}

As a next step, we construct a sequence of networks $G_i$ for the month of May from 1973 to 2013. In Figure~\ref{holanda_time} (A) the empirical expected depth ($m-\hat{D}(G_i)$) for each year is shown. The month of May was selected because in this month there exists four outlying years (1975, 1984, 1994 and 2012) and the latest has occurred recently.  The corresponding networks are shown in Figure~\ref{holanda_time} (B). All four of them have fewer links than the central network which is complete in this case. In order to understand the outlying behavior, in Panel (C) we show the thermal amplitude time series for the six meteorological stations at two normal years (2011 and 2013) and at the outlying year 2012. It can be observed that in normal years all stations present very similar behavior, however at 2012 the most South-West station at the city of Vlissingen (station $e$) shows some departure from this general behavior. The thermal amplitude from station $e$ at may of 2012 has a small dependence with the other stations. This last fact generates a climate network with the station $e$ disconnected with the others stations, as it is shown in Panel B (middle network). Probably, the absence of links of station $e$ at may 2012 is related to particular changes in the air currents from the sea.  Vlissingen is the only of the six cities that has a temperate oceanic climate that is milder than the rest of the Netherlands. It is one of the sunniest cities in the Netherlands, and has the lowest rainfall.  The other five stations are located at Amsterdam, De Bilt, Leeuwarden, Twente, and Zuid-Limburg and are far (at least 20km) from the sea. Although further analysis is needed to confirm this hypothesis, it seems that the the proximity to the sea can be a plausible scenario for explaining the phenomena observed in May of 2012 at Vlissingen. 

\begin{figure}[h]
\centering
\hspace*{-0.5cm}
\includegraphics[angle=0,width=0.8\textwidth]{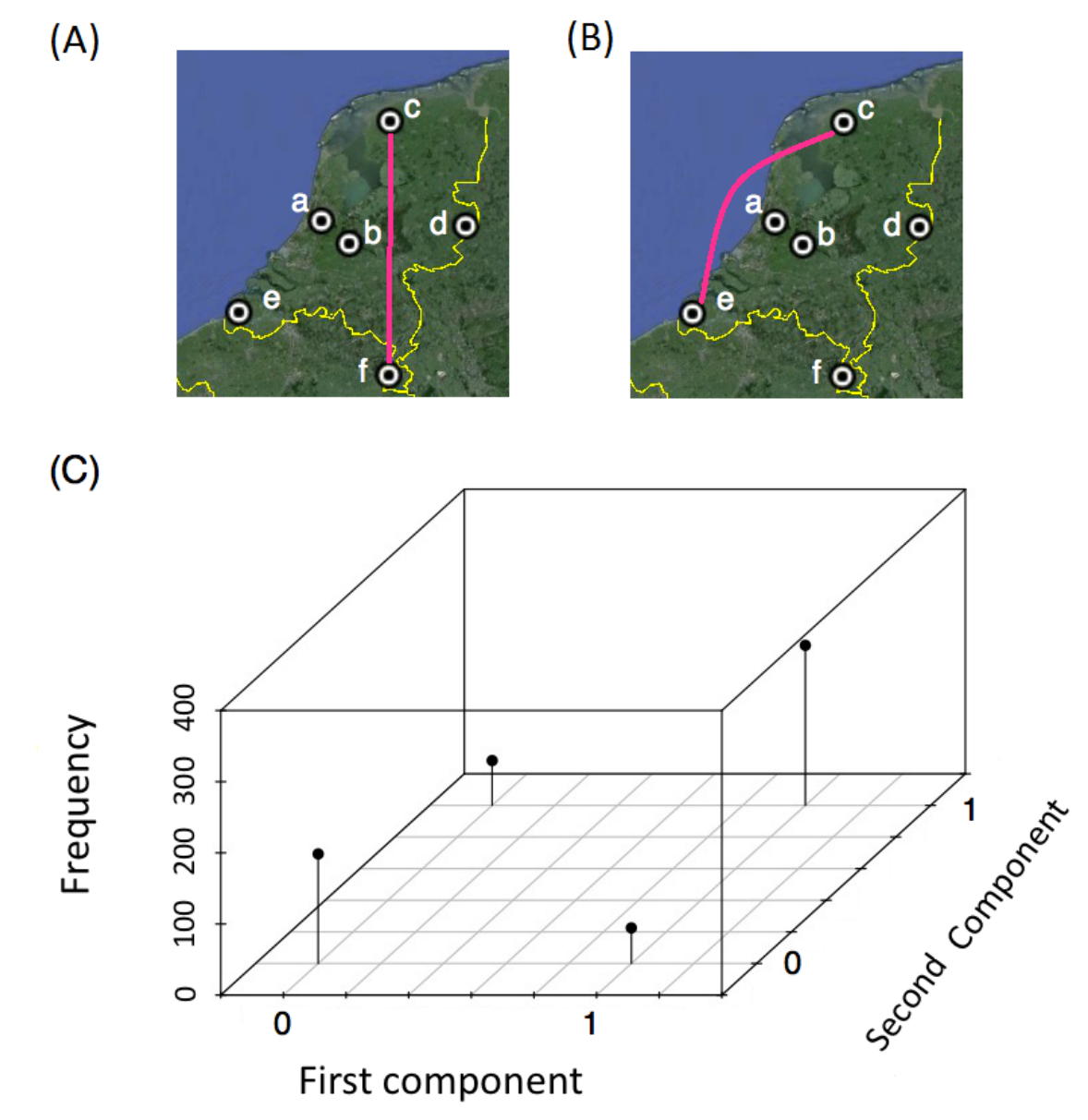}
\caption{(A) First (B) Second principal component. (C) 2D map of the projected data on the first and second principal components.}
\label{compo2}
 \end{figure}

Next, we perform principal component analysis. In this case we study 492 graphs each corresponding to a month of the 41 years in the sample. We look for the interactions between meteorological stations that present large variability. The unique first and second principal components are shown in Fig.~\ref{compo2} (A) and (B) respectively. In this case the variances of the projected data in both principal components are close. Note that there exist a peak at (1,1) on Fig.~\ref{compo2} (C) that reflects the fact that the links $(e,c)$ and $(c,f)$ tend to be present together. Note that these two components links the more distant meteorological stations, i.e. the variability of the ``interaction'' between two stations is greater between distant stations. 

Finally, we perform the test proposed in Section 4 to compare two samples of networks. We compare the sample of networks from the month of January with other months. Each sample has a size of 41 (the numbers of years studied). The network distribution from February or December should be very similar to the one from January because an obvious continuity effect. However, when comparing January to June or July marked differences should appear. We tested the null hypothesis of equal distribution between January and each month of the year, i.e. $\mu_{\text{Jan}} =\mu_{x}$ for each month $x$ using five hundred random projections. Even with a small sample size the test has an excellent performance. 
In particular, extremely small $p$--values were obtained for the months of March through September (see Table~1).

\begin{table}[htp]
\begin{center}
\begin{tabular}{|c|c|c|c|}  \hline
& February & March & April \\ \hline
& 0.035 & $< 1 \times 10^{-50}$ & $< 1\times 10^{-50}$ \\ \hline \hline
May & June & July & August \\ \hline
0 & $< 1 \times 10^{-50}$ & 0 & $< 1\times 10^{-50}$ \\ \hline \hline
September & October & November & December \\ \hline
$<1\times 1^{-50}$ & 0.61 & 0.67 & 1 \\ \hline
\end{tabular}
\end{center}
\caption{P-values for each of the test comparisons ($H_0 :
\mu^{\text{Jan}}_G=\mu_G^{x}$). }
\label{default}
\end{table}%

\section{Discussion}
%==============================
We propose a non parametric framework to study dynamic random networks with a fixed number of distinguishable nodes. Some classical statistical problems such as clustering and principal component analysis are addressed. All the statistics we define have been constructed using a natural distance between networks and its corresponding $L^1$--depth notion. All the results presented here can be easily adapted for another distance. From a theoretical point of view, we believe that the framework we present can be the building block to construct more sophisticated statistical parametric and non parametric techniques. 

One major issue when working with network data is the implementation of theoretical results for large scale networks. While many of the definitions we present involve maximizing over the entire space of networks, we find explicit simple algorithms for the optimization problems which allow to implement our methods for large scale networks and big datasets since the complexity is of order $n^2$ (linear in the number of possible links).

There are lots of possible applications for the results we present. We analyze two real data examples: the first one corresponds to a face--to--face social interaction network evolving in time,  while the second one is an example of the so called correlation networks, for climate data. The results are quite encouraging. It is important to remark that the analysis of functional networks is becoming a standard procedure in the areas of finance~\citep{vespignani,daniel4} and neuroscience~\citep{daniel1,daniel2,sporns}.  

Another very important application in Neuroscience is related to the development of new diagnostic methods based on brain network data. This is directly related to the problem of classification of patients (e.g., high or low-risk to have a particular cognitive disorder) for example from their (fMRI, MEG, or EEG) resting state functional networks.
\section{Acknowledgments}
The authors would like to thank two anonymous reviewers for helpful comments and criticism on earlier versions of the paper. 
%\newpage
%==============================
\vspace{1.5cm}
\appendix
\section{Proofs}
\subsection{Characterization of the central set}

\begin{proof}[Proof of Proposition 1]
We have that the expected distance from a network $H$ to a random network $\mathbf G$ is 
\begin{equation}\label{edistancia}
\Ex{d(\G, H)} = \sum_{G \in \mathcal G} d(G,H)p_G.
\end{equation}
Let $A(G)$ be the adjacency matrix of the network $G$ and $\A$ the adjacency matrix of the random network $\G$. Then expression \eqref{edistancia} can be written as
\begin{align*} 
\sum_{G \in \cG} \sum_{i>j} \vert A(G)_{ij} - A(H)_{ij} \vert p_G 
&= \sum_{i>j} \sum_{G \in \cG} \vert A(G)_{ij} - A(H)_{ij} \vert p_G \\ 
&= \sum_{i>j}\Pr{\A_{ij} \neq A(H)_{ij} } ,
\end{align*}
which is minimized by any network $H$ with adjacency matrix $A(H)_{ij}=1$ if and only if $\Pr{\A_{ij}=1} \geq 1/2$. Moreover, if for all $i,j$
\begin{equation}\label{unico}
\Pr{ \A_{ij}=1} \neq 1/2,
\end{equation}
there is a unique network $S$ that minimizes expression \eqref{edistancia} and the corresponding adjacency matrix satisfies $A(S)_{ij}=1$ if and only if $\Pr{\A_{ij} =1} > 1/2$.

On the other hand, if condition \eqref{unico} does not hold, there are many solutions. The maximal center $L$ is the network whose adjacency matrix fulfills $A(L)_{ij} =1$ when $\Pr{\A_{ij}=1} \geq 1/2$ and the set $\mathcal C$ contains exactly all subnetworks of $L$ for which $S$ is a subnetwork.

The proof for the empirical version is completely analogous.
\end{proof}

%==============================
\subsection{Depth determines measure}

\begin{proof}[Proof of Proposition 2]

Let $\mu = (\mu_1, \ldots, \mu_K)$ and $\nu = (\nu_1, \ldots, \nu_K)$ be two distributions on $\cG$ where $K=2^m$ is the cardinal of the space of networks. For any $H \in \cG$, let $d(H) = (d(H_1,H), \ldots d(H_{K},H))$. Then, the population depth is
$$D_\mu(H) = m - d(H)^T \mu.$$

Therefore, the depth determines the measure if and only if
\begin{equation}\label{coinciden} 
d(H)^T \mu - d(H)^T \nu = 0 \text{ for all $H\in\cG$ implies } \mu = \nu.
\end{equation}

Denote by $F$ the matrix with rows given by $d(H_1), \ldots d(H_{K})$. Then expression (\ref{coinciden}) is equivalent to $F (\mu-\nu) = 0$, having a unique solution. The result follows from the invertibility of distance matrices. This result was initially proved by \citet{xx} and later on by \citet{auer}, who provided an elementary proof, that the only uses the triangle inequality. In particular, our metric is just the $L^1$ distance between the adjacency matrices and the result holds. For the sake of completeness we now state the result in \citet{auer} for distance matrices.

\begin{atheorem}[A]\label{qdistancias}
Let $P_1, \ldots, P_n$ be distinct points in $\mathbb R^k$, and $d_{i,j} = \Vert P_i - P_j \Vert$. If $F_n$ is the distance matrix with entries $d_{i,j}$, then
\begin{itemize}
\item [a)] The $\det F_n$ is positive if $n$ is odd and negative if $n$ is even, in particular $F_n$ is invertible.
\item [b)] The matrix $F_n$ has one positive and $n-1$ negative eigenvalues.
\end{itemize} 
\end{atheorem}

Applying Theorem A to our setup, we get that the matrix $F$ is invertible.
\end{proof}

%==============================

%==============================
\subsection{Convergence of empirical depth}

\begin{proof}[Proof of Theorem 2]
From the Ergodic Theorem we have that $\hat D_\ell(H) \to D(H)$ almost surely as $\ell \to \infty$ for each $H \in \mathcal G$. Since $\mathcal G$ is finite we get uniform convergence.
\end{proof}

\begin{proof}[Proof of Theorem 3]
Recall that  a sequence of random elements $\mathbf X := (\mathbf X_t, t \geq 1)$ is a strong mixing sequence if it fulfills the following condition.
For $ 1 \leq j < \ell \leq \infty$ , let $\mathcal{F}_j^\ell$ denote the $ \sigma$ -field of events generated by the random elements $X_k,\ j \le k \leq \ell\ (k \in {\bf N})$ . For any two $ \sigma$-fields $ \mathcal{A}$ and $ \mathcal{B}$, define
$$
\alpha(\mathcal{A}, \mathcal{B}) := \sup_{A \in \mathcal{A}, B \in \mathcal{B}} \vert \Pr{A \cap B} - \Pr{A} \Pr{B} \vert.
$$
For the given random sequence $\mathbf X$ , for any positive integer $n$, define the dependence coefficient
$$
\alpha(n) = \alpha(\mathbf X,n) := \sup_{j \geq 1}\; \alpha(\mathcal{F}_{1}^j, \mathcal{F}_{j + n}^{\infty}).
$$
The random sequence $\mathbf X$ is said to be ``strongly mixing'', or ``$ \alpha$ -mixing'', if $ \alpha(n) \to 0$ as $ n \to \infty$. This condition was introduced by \citet{rosenblatt56}. By assumption, we have that the sequence of random networks $\{\G_t: t\geq 1 \}$ is a strongly mixing sequence. In order to prove the theorem we use the following result \citep[see for instance][]{peligrad86}.

\begin{atheorem}[B]\label{CLT} 
Let $\{\mathbf X_t: t \geq 1\}$ be a strictly stationary centered $\alpha$--mixing sequence, and let
$\mathbf S_\ell = \sum_{t=1}^\ell \mathbf X_t$. Assume that for some $C>0$
$$
\vert \mathbf X_1 \vert < C \ \mbox{almost surely, and} \ \sum_{n=1}^\infty \alpha(n) < \infty.
$$
Then,
$$
\sigma^2 := \Ex{\mathbf X_1^2} + 2 \sum_{k=2}^\infty \Ex{\mathbf X_1 \mathbf X_k} ,
$$
is absolutely summable. If in addition $\sigma^2 >0$, then $\mathbf S_\ell/\sqrt{\ell}\sigma$ converges weakly to a
standard normal distribution.
\end{atheorem} \bigskip

First observe that
$$
\beta^T \mathbf Z_{\ell}  = \frac{1}{\ell} \sum_{k=1}^{\ell} \mathbf W_k,
\qquad\text{with}\qquad
\mathbf W_k= \sum_{j=1}^{2^m} \beta_j \left( d(H_j, \G_k) - \Ex{d(H_j,\G_1)} \right),
$$
where  $\{\mathbf W_t: t \geq 1 \}$ is a strictly stationary, bounded, centered $\alpha$--mixing sequence, fulfilling $\sum_{n=1}^\infty \alpha(n) < \infty$. On the other hand, we have that
$$
\Ex{\mathbf W_1^2} = \beta^T \Ex{\mathbf Y_1^T \mathbf Y_1} \beta \ \mbox{and}
\ \Ex{\mathbf W_1 \mathbf W_k} = \beta^T \Ex{\mathbf Y_1^T \mathbf Y_k} \beta,
$$
and the result follows from Theorem B.
\end{proof}

%===============================
\subsection{Characterization of principal components}

\begin{proof}[Proof of Proposition 3]
Note that
$$
\Var\left( \frac{|\G\wedge Q|}{|Q|} \right) = \frac{1}{\vert Q \vert^2}\sum_{G \in \mathcal G} \left(\sum_{i>j} c_{ij} A(G)_{ij} A(Q)_{ij} - \sum_{H \in \mathcal G} \sum_{i>j} c_{ij} A(H)_{ij} A(Q)_{ij}p_H \right)^2p_G.
$$

We first consider the case when the network $Q$ has only one link $(k,\ell)$, and find within this family the one that maximizes the objective function. Next we prove that for any other network $Q$ the objective function is bounded by the maximum restricted to the former family. Finally, we show that the principal component space is generated by the one link networks.

Let $Q_1$ such that $A(Q_1)_{k_1\ell_1}=1$, and $0$ otherwise. Also, let $\mathcal G_{Q_1}^+ = \{G\in \mathcal{G}: A(G)_{k_1\ell_1} = 1\}$ and $\mathcal G_{Q_1}^- = \{G\in \mathcal{G}: A(G)_{k_1\ell_1} = 0 \}$. When we search within the one link networks, the objective function reduces to
\begin{align*}
\Var\left( \frac{|\G\wedge Q|}{|Q|} \right) 
&= \frac{1}{c_{k_1\ell_1}^2}\sum_{G \in \mathcal G} \left( c_{k_1\ell_1} A(G)_{k_1\ell_1} - \sum_{H \in \mathcal G} c_{k_1\ell_1} A(H)_{k_1\ell_1} p_H	\right)^2 p_G \\
&=\sum_{G \in \mathcal G}  \left( A(G)_{k_1\ell_1} -\Pr{\A_{k_1\ell_1}=1} \right)^2 p_G \\
&=\sum_{G \in \mathcal G^+}   \left( 1- \Pr{\A_{k_1\ell_1}=1} \right)^2 p_G + \sum_{G \in \mathcal G^-} \Pr{\A_{k_1\ell_1}=1} ^2 p_G \\
&= ( 1- \Pr{\A_{k_1\ell_1}=1} )^2 \Pr{\A_{k_1\ell_1}=1} + \Pr{\A_{k_1\ell_1}=1} ^2 \Pr{\A_{k_1\ell_1}=0} \\
&= \Pr{\A_{k_1\ell_1}=1} (1- \Pr{\A_{k_1\ell_1}=1} ),
\end{align*}
and the solution is the one link graph for which $\Pr{\A_{k_1\ell_1}=1} $ is closest to $1/2$.

In the general case, we want to find $Q$ that maximizes
\begin{align*}
\Var\left( \frac{|\G\wedge Q|}{|Q|} \right) 
&= \frac{1}{\vert Q \vert^2} \sum_{G \in \mathcal G} \left( \sum_{(i,j)\in Q} c_{ij} A(G)_{ij} - \sum_{H \in \mathcal G} \sum_{(i,j)\in Q} c_{ij} A(H)_{ij}  p_H \right)^2 p_G \\
&= \sum_{G \in \mathcal G} \ \sum_{(i,j)\in Q} w_{ij} \left(A(G)_{ij} - \Pr{\A_{ij}=1} \right)^2 p_G, %\label{equiv}
\end{align*}
where $w_{ij}= c_{ij} / \sum_{(p,q)\in Q} c_{pq}$. Now, since the weights $w_{ij}$ add to one, we have
\begin{align*}
\Var\left( \frac{|\G\wedge Q|}{|Q|} \right)  
&\leq \sum_{(i,j)\in Q} w_{ij}\; \max_{(i,j)}\; \Pr{\A_{ij}=1} \left(1 - \Pr{\A_{ij}=1} \right) \\
&= \max_{(i,j)}\; \Pr{\A_{ij}=1} \left(1-\Pr{\A_{ij}=1} \right),
\end{align*}
which corresponds to the one link optimum. 
\ \

If there exist a unique one link graph ($Q_1$) that verifies $\Pr{\A_{k_1\ell_1}=1}$ is closest to $1/2$, then the \textit{principal component space} is generated just by $Q_1$, i.e $\cS_1 =\mathcal G_{Q_1}^+$.  If there exist multiple one link graphs, $Q_1, Q_2, \dots, Q_p$, that minimize $|\Pr{\A_{k\ell}=1}-1/2|$. then the \textit{principal component space} is $\cS_1 = \cup_{i=1}^p \mathcal G_{Q_i}^+$. The second \textit{principal component space} verifies the same. In this case the maximization of the variance  is over  $\{G\in \mathcal{G}: G \notin \cS_1 \}$. Analogous for the rest of the components, for example for finding the $k-esima$ \textit{principal component space} just maximizes the variance over $\{G\in \mathcal{G}: G \notin \cS_1, G \notin \cS_2, G \notin \cS_{k-1} \}.$

\end{proof}	

%==============================
\subsection{Consistency of Principal Components}

\begin{proof}[Proof of Proposition 4]

For each $Q \in \mathcal G$ from the Ergodic Theorem we have the following.
\begin{itemize}
\item [a)] $\displaystyle \Lambda_{\ell} (Q) = \frac{1}{\ell} \sum_{k=1}^{\ell} \vert G_k \wedge Q \vert \to \Ex{ \vert\G \wedge Q\vert } $,
almost surely as $\ell \to \infty$.

\item [b)]  
$\aligned[t]
\Delta_{\ell}(Q) &= \widehat{\Var} \left( \frac{\vert \mathbf G \wedge Q \vert}{Q}\right)  &\\
&= \frac{1}{\ell} \sum_{k=1}^{\ell} \left(\frac{ \vert G_k \wedge Q \vert - \Lambda_{\ell}(Q)}{\vert Q \vert}\right)^2 &\\ 
&= \frac{1}{\ell} \sum_{k=1}^{\ell} \frac{ \vert G_k \wedge Q\vert^2}{\vert Q \vert^2} + \frac{\Delta_{\ell}^2(Q)}{\vert Q \vert^2} 
-  \frac{2}{\ell} \sum_{k=1}^{\ell} \frac{ \vert G_k \wedge Q\vert}{\vert Q \vert^2} \Lambda_{\ell}(Q), &
\endaligned$\par
which converges almost surely to
\begin{equation} \label{var}
\Ex{ \frac{\vert \G \wedge Q \vert^2}{\vert Q \vert^2}} 
+ \Ex{ \frac{\vert \G \wedge Q \vert}{\vert Q \vert^2}} ^2 
- 2 \Ex{ \frac{\vert \G \wedge Q \vert}{\vert Q \vert^2}} ^2
= \Var\left( \frac{\vert \G \wedge Q \vert}{\vert Q \vert}\right).
\end{equation}
\item [c)] Since the space $\mathcal G$ is finite expression (\ref{var}) entails that $\hat{\mathcal Q}_1 \to \mathcal Q_1$ almost surely, i.e., $ \hat{\mathcal Q}_1 = \mathcal Q_1$ for $\ell$ large enough almost surely, which entails that the principal components converge because the geodesics coincide eventually.
\end{itemize}
For the next principal component the proof is analogous.
\end{proof}

%==============================
\bigskip
\bibliographystyle{spbasic} 
\bibliography{refs}

\end{document}